\documentclass{article}
\usepackage{jinst}
\usepackage{lineno}

\title{Energy reconstruction in a liquid argon calorimeter cell using convolutional neural networks}
\author[a]{L.~Polson,}
\author[b]{L.~Kurchaninov}
\author[a]{and M.~Lefebvre}
\affiliation[a]{University of Victoria,\\
3800 Finnerty Road, Victoria, Canada}
\affiliation[b]{TRIUMF,\\
4004 Wesbrook Mall, Vancouver, Canada}
\emailAdd{lpolson@uvic.ca}

\abstract{The liquid argon ionization current in a sampling calorimeter cell can be analyzed to determine the energy of detected particles. In practice, experimental artifacts such as pileup and electronic noise make the inference of energy from current a difficult process.  The beam intensity of the Large Hadron Collider will be significantly increased during the Phase-II long shut-down of 2025-2027. Signal processing techniques that are used to extract the energy of detected particles in the ATLAS detector will suffer a significant loss in performance under these conditions. This paper compares the presently used optimal filter technique to convolutional neural networks for energy reconstruction in the ATLAS liquid argon hadronic end cap calorimeter. In particular, it is shown that convolutional neural networks trained with an appropriately tuned and novel loss function are able to outperform the optimal filter technique.}


\begin{document}
\bibliographystyle{unsrt}
\maketitle

\section{Introduction}

The inference of energy from measured current in ionization detectors is often a non-trivial process, particularly in the regime of high energy particle physics. The conditions in the liquid argon (LAr) calorimeter of the ATLAS detector~\cite{atlas,LHC,lar_cal} expected after the high luminosity (HL) upgrade to the Large Hadron Collider (LHC)~\cite{LHC,hl_lhc} are such that the precision of energy measurements will suffer a loss in performance. This paper seeks to simulate the conditions of the HL-LHC and the electronic response of a particular detector cell in the hadronic end cap (HEC) subsystem of the ATLAS LAr Calorimeter. The presently used algorithm for energy reconstruction from measured signal, known as the optimal filter (OF)~\cite{OF}, will be compared to convolutional neural network (CNN) architectures. It will be shown that a model architecture trained with an appropriate loss function on simulated data outperforms the optimal filter in relevant metrics. 

\section{Simulation of Current in a HEC Calorimeter Cell}

The total energy deposited in a HEC detector cell $E$ is given by the sum of two different components: pileup $p$, which includes deposited energy from multiple low-energy collisions and is proportional to the luminosity of the LHC, and signal $S$, which is representative of particular and rare processes. The energy can be expressed as a function of discrete time $t$, with 25~ns or 1~bunch crossing (BC) spacing:

\begin{equation}
    E(t) = S(t) + p(t)\, .
\end{equation}

In this study, $p(t)$ is simulated according to~\cite{hec110}, where the total pileup energy at a given BC is the sum of energies from many minimum bias events. The probability density function for the energy of a single minimum bias event is given by

\begin{equation}
    P(E) = q\delta(E) + (1-q)\left( F_1 e^{-\sqrt{E/E_1}} + F_2 e^{-\sqrt{E/E_2}} \right) \, , \label{pileup_pde}
\end{equation} 
where $\delta(E)$ is the Dirac delta function and the expression is written such that $q$ is the probability of zero energy deposition.
The parameters used are $F_1=1.721~\text{GeV}^{-1}$, $E_1=1.012~\text{GeV}$, $F_2=0.0226~\text{GeV}^{-1}$, $E_2=1.012~\text{GeV}$, and $q=0.8314$;
they correspond to the expected pileup conditions at $\eta = 2.25$ in the first layer of the HEC as obtained through simulations in~\cite{hec110} and were provided by one of the authors of this note\footnote{The statistical uncertainty of these parameters is small and sufficient to ensure an uncertainty on $P(E)$ that is smaller than the uncertainty of the electronics chain modeling, which is at the level of 2\%.}.
The number of minimum bias events per BC used for simulation is a Poisson random variable with mean $\left< \mu \right> = 200$ -- the expected value at the HL-LHC~\cite{hl_lhc}. These energies are summed together at each BC time $t$ to obtain the total pileup $p(t)$.

$S(t)$ is simulated such that consecutive ionization pulses are non-overlapping\footnote{Consecutive ionization pulses induced by $S(t)$. In general, ionization pulses from pileup $p(t)$ will be overlapping.}; specifically, the spacing between non-zero values is a random variable $\Gamma \sim \mathcal{U}(30~\text{BC},50~\text{BC})$ where $\mathcal{U}(a,b)$ is the uniform distribution between values $a$ and $b$. The distribution of energies during signal injection is given by 

\begin{equation}
    f(s) = \frac{1}{3} \Big[ \mathcal{U}(0~\text{GeV}, 0.3~\text{GeV}) + \mathcal{U}(0.3~\text{GeV}, 5~\text{GeV}) + \mathcal{U}(5~\text{GeV}, 50~\text{GeV}) \Big]\, ,  \label{eq:signal_dist}
\end{equation}
designed to consider relevant signal energies for the HL-LHC.

The total energy $E$ deposited in a HEC detector cell at a given BC produces a triangular-shaped ionization drift current pulse which, upon convolution with the electronics chain response, results in the measurable current $X(t)$.  The electronics chain is described in~\cite{BAN2006158, 109}, and features a bipolar shaper. Sample time series of $X(t)$ and $S(t)$ are shown in the left and middle subplot of Figure~\ref{fig:of_prelim}, respectively.

\section{Energy Reconstruction Using the Optimal Filter Technique}

The signal energy is obtained by inferring $S(t)$ from $X(t)$ and is presently obtained using the OF~\cite{OF}, which produces an estimator $\hat{S}_{\text{OF}}(t)$ given by a weighted sum on $X(t)$ with $N$ coefficients $a_i$:

\begin{equation}
    \hat{S}_{\text{OF}}(t+N) = \sum_{i=1}^N a_i X(t+i) \, . \label{eq:opt}
\end{equation} 

\begin{figure}
\begin{center}
\includegraphics[width=16cm]{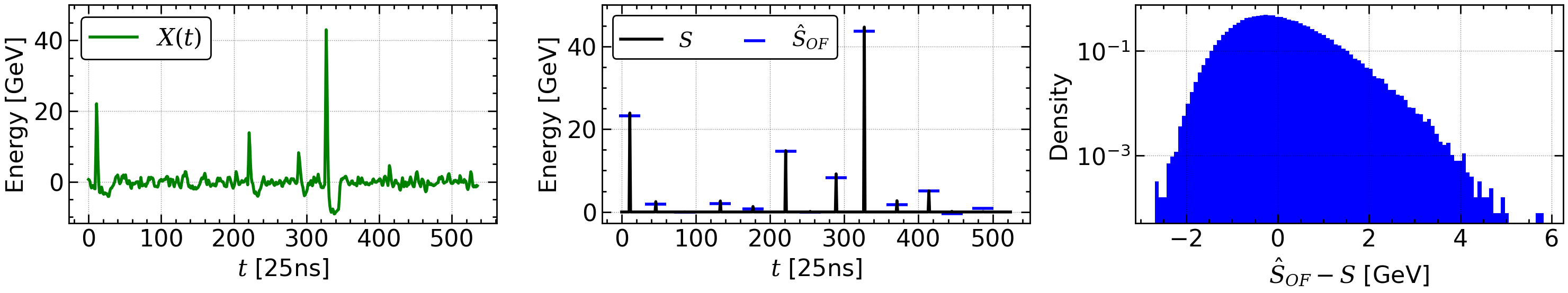}
\end{center}
\caption{Simulated current $X(t)$ representative of the ATLAS experiment [left], signal and optimal filter estimation of the signal [middle] and histogram of residuals $S-\hat{S}_{\text{OF}}$ for all 30 million BCs [right]. For display purposes, $X(t)$, $S$, and $\hat{S}_{OF}$ are aligned so that the peaks coincide.}
\label{fig:of_prelim} 
\end{figure}

All time series used have a length of 30 million BCs. $N=25$ coefficients are used in the OF; previous studies have shown there is no significant performance gain with more coefficients~\cite{thesis}. Shown in Figure~\ref{fig:of_prelim} is a sample of the measurable current $X(t)$ [left], the signal $S(t)$ and optimal filter predictions $\hat{S}_{OF}$ when signal is injected [middle], and a histogram of the residuals between the true and predicted signals for the entire time series [right].

\begin{figure}
\begin{center}
\includegraphics[width=16cm]{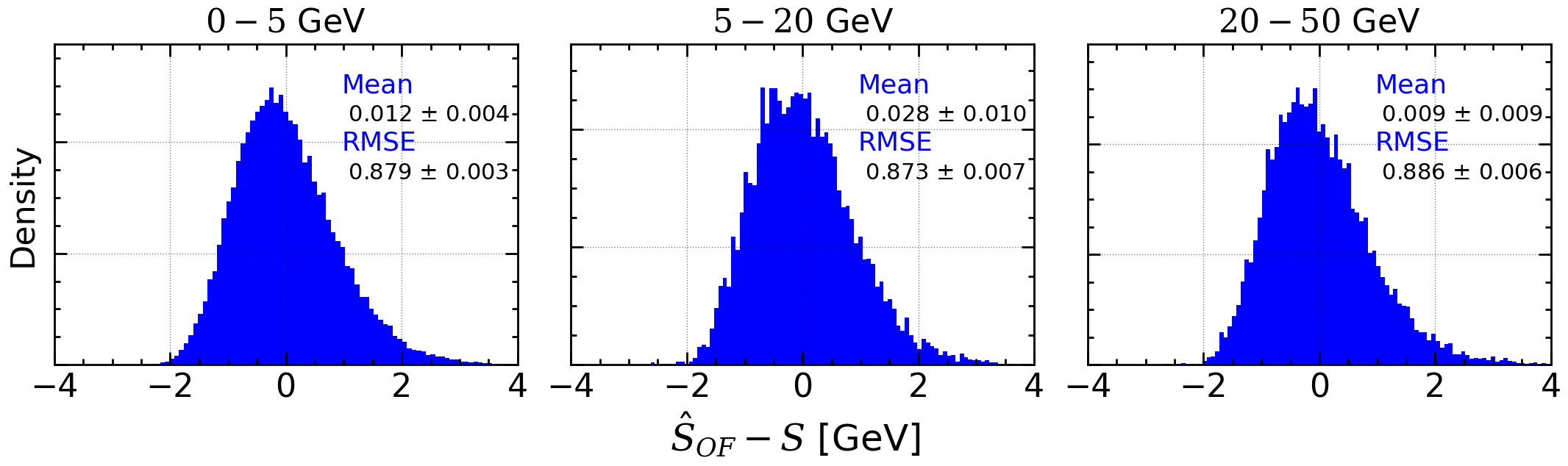}
\end{center}
\caption{Histograms of $\hat{S}_{OF}-S$ for different signal magnitudes $S$, specified by the title of each subplot. The mean and RMSE are shown on each subplot.}
\label{fig:of_regions} 
\end{figure}

A key feature of the OF is its ability to estimate $S(t)$ with little bias in separate energy regions: this is apparent in Figure~\ref{fig:of_regions}, where residuals are separated based on signal magnitude. Such an estimator is said to have no energy range bias; the extent to which an estimator $\hat{S}$ makes locally biased predictions can be quantified by the following metric

\begin{equation}
    B_L(S, \hat{S}) =  \sum_{j=1}^n \left| \frac{1}{N_j}\sum_{S \in P_j} \left(\hat{S}(\theta) - S\right) \right| \, ,
\end{equation}
where $\theta$ represents the parameters of the estimator, $P_j$ is an interval of energies and $N_j$ is the number of events that lie in the interval. Desirable features of any estimator $\hat{S}$ are small $B_L$ and RMSE values. The $P_j$'s used are given by Table~\ref{tab:sig_partition}.

\begin{table}
    \centering
    \begin{tabular}{|c|c|c|c|c|c|c|c|}
    \hline
         Subset & $P_1$ & $P_2$ & $P_3$ & $P_4$ & $P_5$ & $P_6$ & $P_7$\\ \hline
         Values [GeV] & 0-0.3 & 0.3-1 & 1-2 & 2-5 & 5-10 & 10-20 & 20-50\\
         \hline
    \end{tabular}
    \caption{Signal partitions.}
    \label{tab:sig_partition}
\end{table}

\section{Energy Reconstruction Using a CNN}

This section provides evidence that a CNN~\cite{cnn} architecture may outperform the OF for energy reconstruction in a LAr detector cell. A CNN architecture is chosen because it has been shown to be competitive with other types of Machine Learning architectures~\cite{wavenet}, it is relatively fast to train on systems with a graphical processing unit, and it could likely be implemented on field programmable gate arrays (FPGAs)~\cite{chep} that will be deployed in ATLAS in the Phase-II upgrade~\cite{phase2}. All models considered have less than 100 parameters in view of FPGA implementation, though the feasibility of implementation is not guaranteed by this constraint.

Models used were based on a WaveNet architecture~\cite{wavenet} with relatively few layers and filters, and small kernel sizes. A single convolutional layer with a filter size of 3 was used as the final layer for each model. The filter depth, which is defined as the number of input data points used to make a prediction for a single output data point, is dependent on the model hyperparameters. Similar to the OF, the models used were not causal and were able to use 11 future BCs when making predictions. All models were trained for 2000 epochs. Regularization and cross validation were not considered necessary; models did not overfit as the size of the training data was large and the models contained few parameters. The models were trained on 80\% of the data and evaluated on the other 20\%.

\subsection{Using an RMSE Loss Function}

\begin{figure}
\begin{center}
\includegraphics[width=16cm]{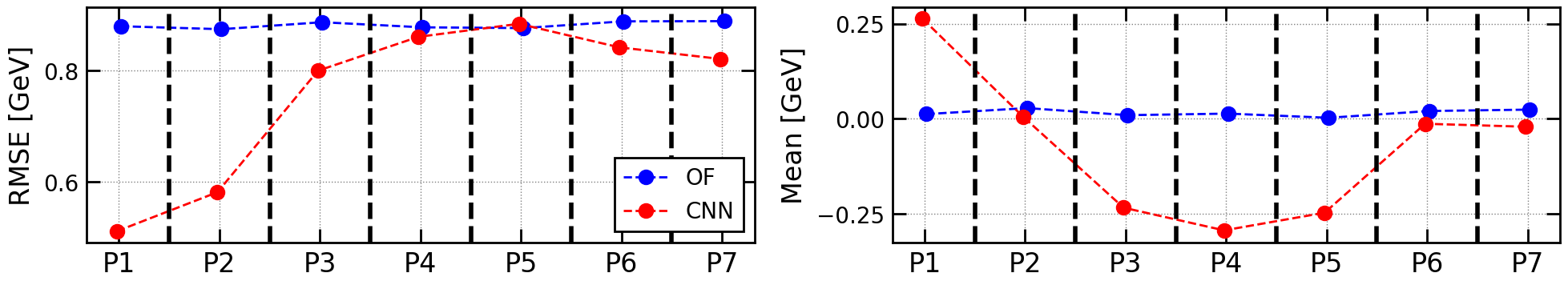}
\end{center}
\caption{RMSE and Mean of residuals $\hat{S}-S$ for the top model architecture of Table~\ref{tab:hyper} and for each of the subsets $P_j$ in Table~\ref{tab:sig_partition}. }
\label{fig:CNN_results1_reduced}
\end{figure}

\begin{figure}
\begin{center}
\includegraphics[width=16cm]{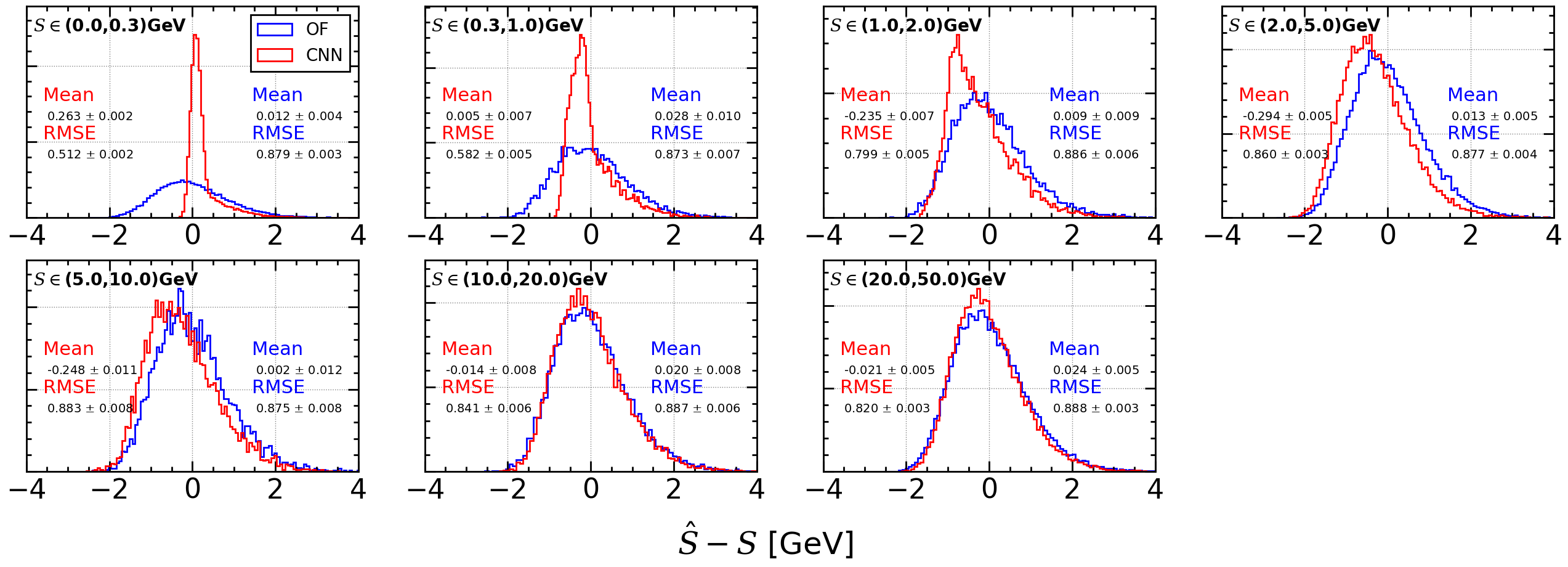}
\end{center}
\caption{Histograms of $\hat{S}-S$ for models trained with a standard RMSE loss function, for each of the subsets $P_j$ in Table~\ref{tab:sig_partition}.}
\label{fig:CNN_results1}
\end{figure}

Table~\ref{tab:hyper} shows the results of a hyperparameter search where models are trained using an RMSE loss function and a Nadam variation of gradient descent~\cite{grad_descent}. All models have less than 100 parameters. While the CNNs tend to make predictions with lower RMSE than those of the OF, the large $B_L$ values are indicative of large local biases in CNN predictions. This is visible, for example, in the predictions of the CNN model architecture with the lowest RMSE score, shown in Figure~\ref{fig:CNN_results1_reduced}. The CNN is able to make predictions with a lower RMSE than the OF, especially at lower energies, but it also tends to overpredict small energies and underpredict intermediate energies. Corresponding histograms of residuals $\hat{S}-S$ in each region $P_j$ are shown in  Figure~\ref{fig:CNN_results1}. Compared to the OF, over all reconstructed signals $\hat{S}$, the RMSE obtained with the CNN is $(17.2 \pm 0.2)$\% lower.

\begin{table}
  \scriptsize
  \centering
  \renewcommand{\arraystretch}{1.2}
  \begin{tabular}{|c|c|c|c|c|c|c|}
    \hline
    \multicolumn{4}{|c|}{\textbf{Hyperparameters}} & \multicolumn{3}{c|}{\textbf{Results}}\\

     Dilations & Filters & Kernel size & Filter Depth & Rank & RMSE [GeV] & $B_L$ [GeV]\\
    \hline
     (1,1) &      3 &     7 & 13&        1 &  0.7294$\pm$0.0002 &      1.117$\pm$0.015 \\
     (1, 1, 2, 2) &      2 &     5 & 25 &      2 &  0.7300$\pm$0.0004 &      1.118$\pm$0.007 \\
     (1, 1, 1, 2) &      2 &     5 &  21 &    3 &  0.7308$\pm$0.0005 &      1.120$\pm$0.009 \\
     (1, 2) &      3 &     7 & 19&        4 &  0.7318$\pm$0.0006 &      1.080$\pm$0.020 \\
     (1, 1, 2) &      2 &     7 & 25 &         5 &  0.7321$\pm$0.0003&      1.131$\pm$0.005 \\
     (1, 3) &      3 &     7 & 25&          6 &   0.7323$\pm$0.0007 &      1.098$\pm$0.017 \\
     (1, 2) &      3 &     5 & 13&         7 &  0.7324$\pm$0.0008 &      1.067$\pm$0.014 \\
     (1, 3) &      3 &     5 & 17&        8 &   0.7341$\pm$0.0004 &      1.081$\pm$0.006 \\
     (1, 1, 3) &      2 &     5 & 21&       9 &   0.7357$\pm$0.0002 &      1.075$\pm$0.006\\
     (1, 3) &      4 &     3 & 9 &        11 &   1.0678$\pm$0.0014 &      2.022$\pm$0.017\\
     \hline
     \multicolumn{4}{|c|}{\textbf{Optimal Filter}}&  10 & 0.8810$\pm$0.0016 & 0.108$\pm$0.021\\
    \hline
  \end{tabular}
  \caption{Hyperparameter search: the ranking is based on the RMSE metric. The OF metrics are shown in the bottom row.}
  \label{tab:hyper}
\end{table}

\subsection{Using an Alternative Loss Function}

\begin{figure}
\begin{center}
\includegraphics[width=13cm]{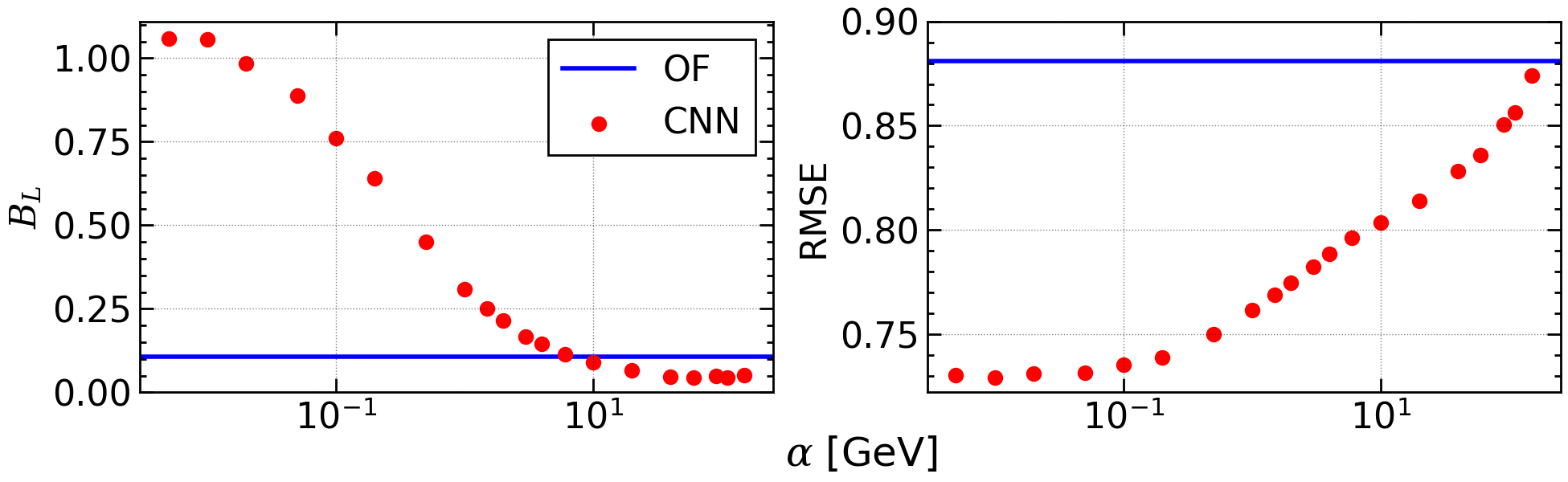}
\end{center}
\caption{RMSE and $B_L$ of $S$ and $\hat{S}$ for different values of $\alpha$. The model hyperparameter configuration is the top configuration from Table~\ref{tab:hyper}. The blue lines represent the optimal filter metrics.}
\label{fig:alpha_search}
\end{figure}

 To prevent CNNs from being optimized in such a way that they make locally biased predictions, the following loss function is proposed

\begin{equation} 
    L(\theta) = \text{RMSE}(\hat{S}(\theta), S) + \alpha \sum_{j=1}^n \left( \frac{1}{N_j}\sum_{S \in P_j} (\hat{S}(\theta) - S) \right)^2  \, ,\label{eq:new_loss}
\end{equation}
where RMSE is the standard RMSE loss function and $\alpha$ is an adjustable constant. In Figure~\ref{fig:alpha_search}, it is shown that a proper adjustment of $\alpha$ results in a CNN predictor with smaller RMSE and $B_L$ than the OF. For small $\alpha$, the loss function is dominated by the RMSE and the model's RMSE and $B_L$ are consistent with that seen in Table~\ref{tab:hyper}. As $\alpha$ increases, locally biased predictions become increasingly penalized, and the model correspondingly has a lower $B_L$ value after training. The trade-off, however, is a slight reduction in the RMSE performance. A tunable $\alpha$ permits an adjustable trade-off between RMSE and $B_L$ metrics; this could potentially be used to minimize net statistical uncertainty in the ATLAS experiment. Hyperparameters were not re-tuned for each value of $\alpha$; it thus may be the case that a different optimal hyperparameter configuration exists for each value of $\alpha$.

\begin{figure}
\begin{center}
\includegraphics[width=16cm]{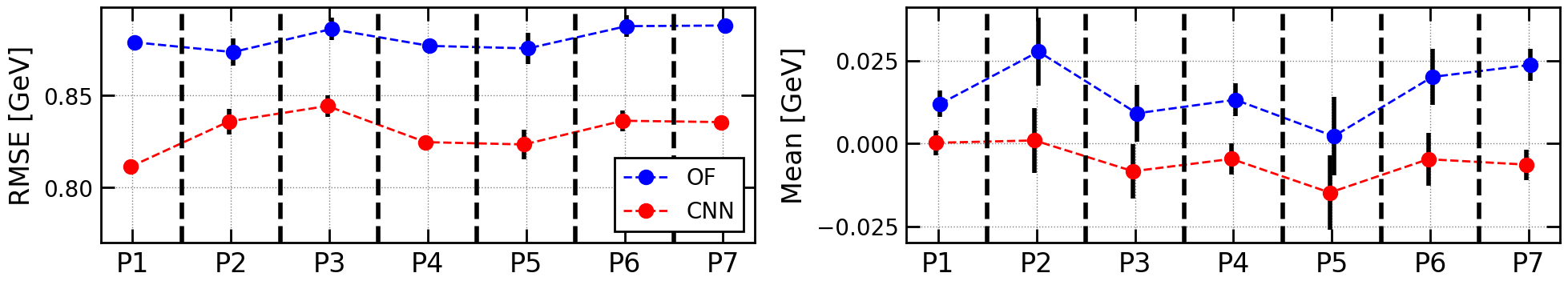}
\end{center}
\caption{RMSE and Mean of residuals $\hat{S}-S$ for the optimal filter and top model architecture in Table~\ref{tab:hyper} trained with $\alpha=40$ in Equation~\ref{eq:new_loss} for each of the subsets $P_j$ in Table~\ref{tab:sig_partition}. }
\label{fig:CNN_results2_reduced}
\end{figure}

\begin{figure}
\begin{center}
\includegraphics[width=16cm]{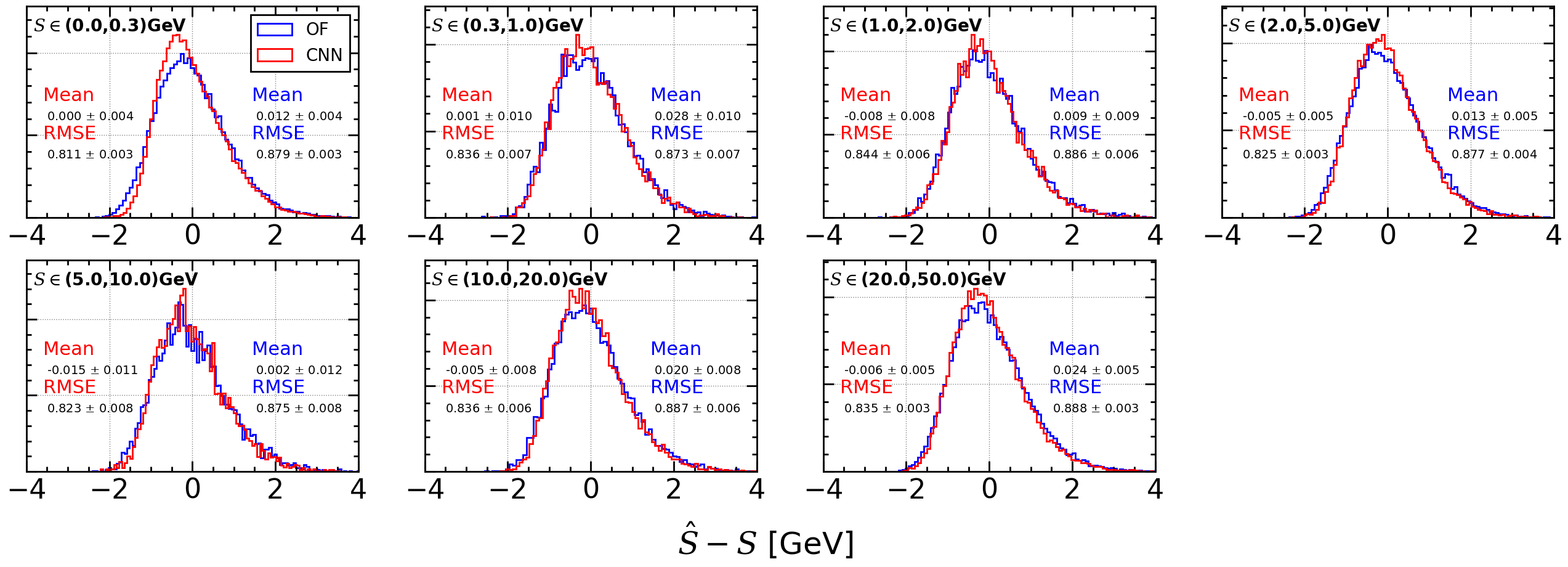}
\end{center}
\caption{Histograms of $\hat{S}-S$ for models trained with the loss function given by Equation~\ref{eq:new_loss} with $\alpha=40$, for each of the subsets $P_j$ in Table~\ref{tab:sig_partition}.}
\label{fig:CNN_results2}
\end{figure}

The mean and RMSE of $\hat{S}-S$ for the top model architecture trained with the value $\alpha=40$ are shown in Figure~\ref{fig:CNN_results2_reduced}; contrasted with Figure~\ref{fig:CNN_results1_reduced}, it is apparent that the new loss function successfully prevents the CNN model from making locally biased predictions.  Corresponding histograms of residuals $\hat{S}-S$ in each region $P_j$ are shown in Figure~\ref{fig:CNN_results2}. Compared to the OF, over all reconstructed signals $\hat{S}$, the RMSE obtained with the CNN is $(6.0 \pm 0.2)$\% lower.

\begin{figure}
\begin{center}
\includegraphics[width=10cm]{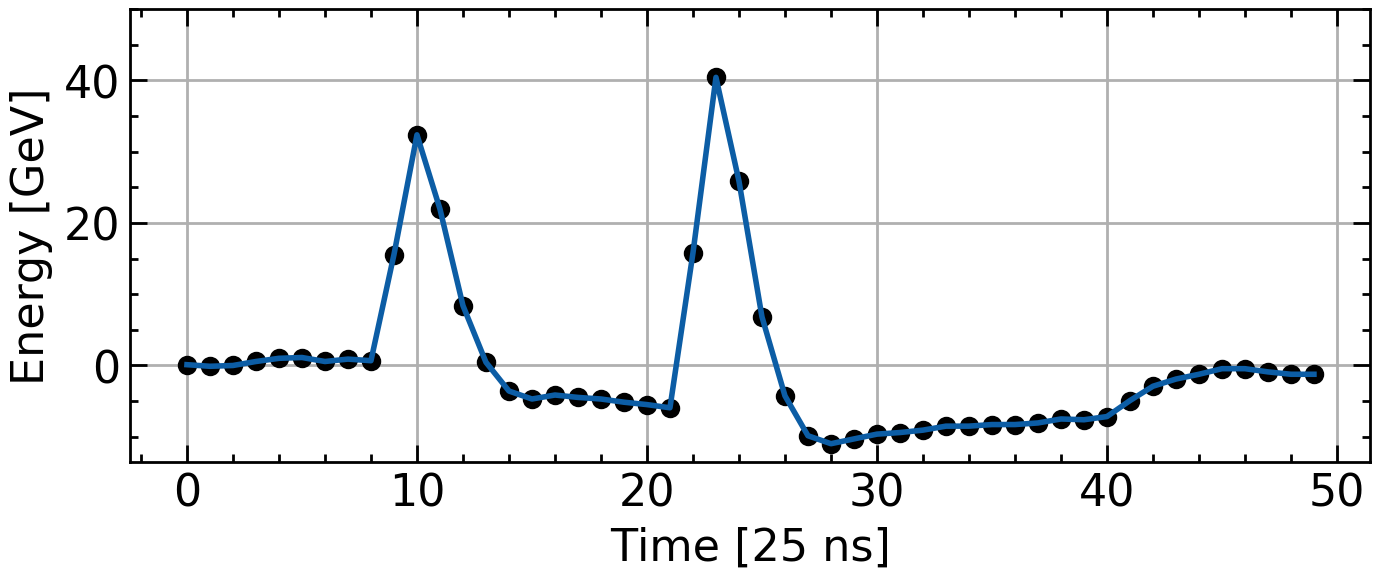}
\end{center}
\caption{Two overlapping ionization pulses from signal events. Black dots represent data and blue lines represent the linear interpolation between data points.}
\label{fig:overlap}
\end{figure}

\begin{figure}
\begin{center}
\includegraphics[width=16cm]{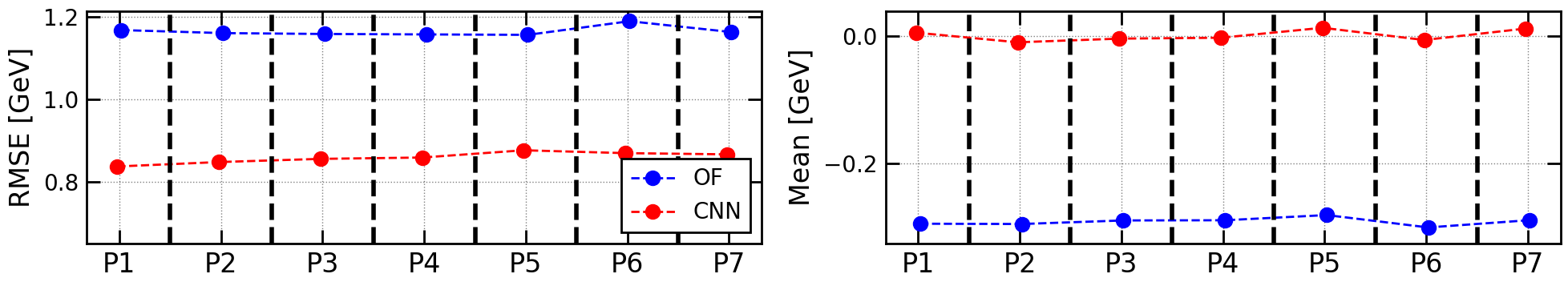}
\end{center}
\caption{RMSE and Mean of $\hat{S}-S$ for the optimal filter and the top model architecture in Table~\ref{tab:hyper} trained with $\alpha=40$ in Equation~\ref{eq:new_loss} for each of the subsets $P_j$ in Table~\ref{tab:sig_partition}.  The data set consisted of signal hits close-by in time.}
\label{fig:CNN_results3_reduced}
\end{figure}

\begin{figure} 
\begin{center}
\includegraphics[width=16cm]{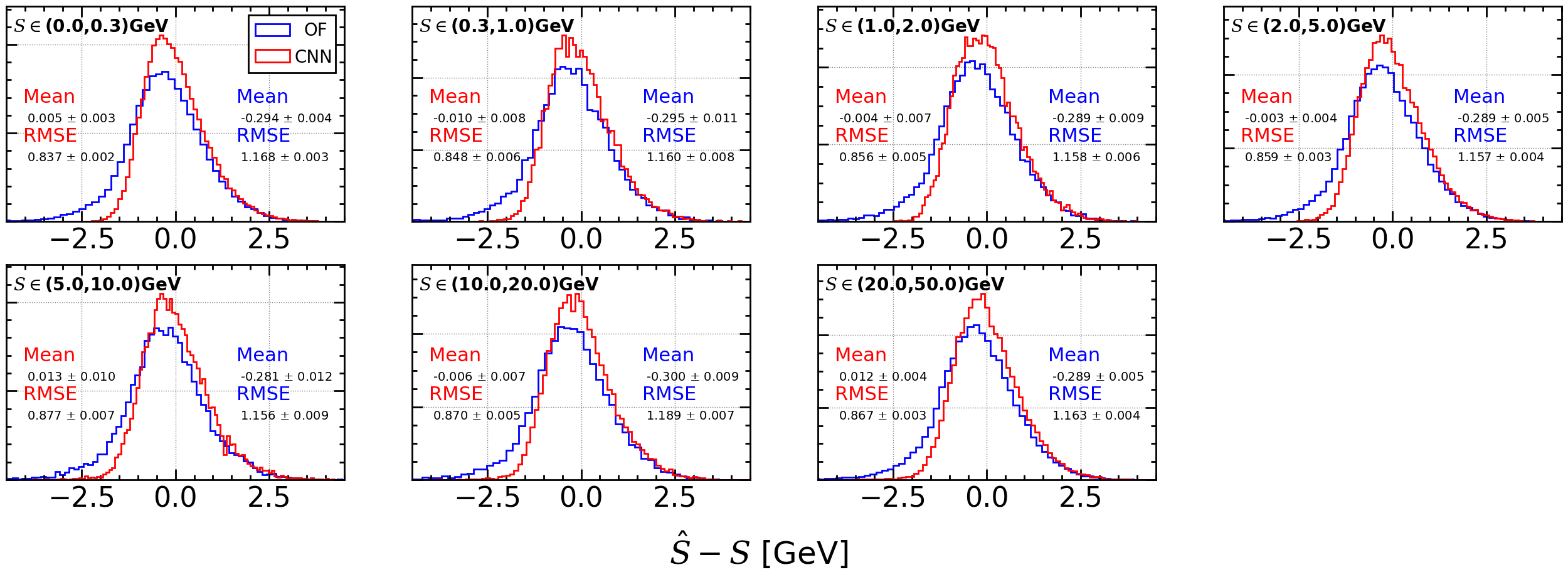}
\end{center}
\caption{Histograms of $\hat{S}-S$ for models trained with the loss function given by Equation~\ref{eq:new_loss} with $\alpha=40$, for each of the subsets $P_j$ in Table~\ref{tab:sig_partition}. Signal is simulated here such that consecutive ionization pulses occasionally overlap.}
\label{fig:CNN_results3}
\end{figure}

The optimal hyperparameter configuration and loss function can additionally be used to train a model on a dataset where the signal $S$ from Figure 1 is spaced sufficiently close together that overlapping ionization pulses are obtained from consecutive signal events. This is simulated by using $\Gamma \sim \mathcal{U}(3~\text{BC},50~\text{BC})$; a sample of $X(t)$ obtained using this $\Gamma$ is shown in Figure~\ref{fig:overlap}. The performance of the CNN is significantly greater than the OF in this regime; this is shown in Figure~\ref{fig:CNN_results3_reduced}.  Corresponding histograms of residuals $\hat{S}-S$ in each region $P_j$ are shown in Figure~\ref{fig:CNN_results3}. Compared to the OF, over all reconstructed signals $\hat{S}$, the RMSE obtained with the CNN is $(26.2 \pm 0.2)$\% lower. In this regime, the OF makes locally biased predictions in all intervals but the CNN does not. In addition, the relative difference between the RMSE of the CNN and OF increases.

\section{Conclusion}

A convolutional neural network (CNN) architecture was explored as an alternative to the optimal filter (OF) energy reconstruction algorithm in a HEC ATLAS liquid argon calorimeter cell. Signal was simulated according to expected pileup conditions in ATLAS at the HL-LHC without overlapping signal ionization pulses. A standard RMSE loss function was explored as a first option for training CNNs; while the CNN saw a reduction in RMSE of $(17.2 \pm 0.2)$\% compared to the OF, the CNN also developed locally biased predictions in different energy ranges. A novel loss function for training was developed to address this bias. This loss function eliminated energy range biases and the CNN still outperformed the OF in RMSE by $(6.0 \pm 0.2)$\%. Signal was then simulated such that consecutive events could lead to overlapping ionization pulses; in this case, the CNN had the greatest performance improvement over the OF, with a $(26.2\pm0.2)$\% decrease in RMSE and no energy range bias. With the new loss function, a CNN architecture outperforms the OF in RMSE while eliminating energy range biases.

\section*{Acknowledgments}
The authors would like to thank Dr Steffen St\"{a}rz and Alessandro Ambler for fruitful discussions.
Support from the Natural Sciences and Engineering Research Council of Canada, Compute Canada, and CMC Microsystems is greatfully acknowledged.



\end{document}